\documentclass{nature}
\usepackage{graphicx}
\makeatletter
\let\saved@includegraphics\includegraphics
\AtBeginDocument{\let\includegraphics\saved@includegraphics}
\renewenvironment*{figure}{\@float{figure}}{\end@float}
\renewenvironment*{table}{\@float{table}}{\end@float}
\makeatother
\usepackage[utf8]{inputenc}
\usepackage{microtype}
\usepackage{graphicx}
\usepackage{subfigure}
\usepackage{booktabs} % for professional tables
\usepackage{chngcntr}
\usepackage{amsmath}
\usepackage{xcolor}
\usepackage{amssymb}
\usepackage{hyperref}
\usepackage{stackengine}
\usepackage{authblk}
\usepackage{soul}
\usepackage{siunitx}
\usepackage[version=4]{mhchem}

\newcommand{\tx}[1]{#1}

\newcommand{\mh}{{MLP}}
\newcommand{\mhsim}{{MLP+}}
\newcommand{\gn}{{AGN}}
\newcommand{\gnsimm}{{AGN+}}

\newcommand{\compilewithappendix}{true}
\newcommand{\supautoref}[2]{\ifthenelse{\equal{\compilewithappendix}{true} }{\autoref{#1}}{#2}}
\newcommand{\supref}[2]{\ifthenelse{\equal{\compilewithappendix}{true} }{\ref{#1}}{#2}}

\newcommand{\supplementary}{Supplementary Information}

\usepackage{bibunits}

\newcommand{\beginsupplement}{%
        \setcounter{table}{0}
        \setcounter{figure}{0}
        \renewcommand{\tablename}{Supplementary Table}
        \renewcommand{\figurename}{Supplementary Figure}
        \renewcommand{\thesection}{Supplementary Section \Alph{section}}%
        \def\figureautorefname{Supplementary Figure}
        \def\tableautorefname{Supplementary Table}
     }

\begin{document}

\begin{bibunit}[naturemag]

\title{Atomistic graph networks for experimental materials \\property prediction}
\author[1,2*]{Tian Xie}
\author[1*]{Victor Bapst}
\author[1]{Alexander L. Gaunt}
\author[1]{\mbox{Annette Obika}}
\author[1]{\mbox{Trevor Back}}
\author[1]{\mbox{Demis Hassabis}}
\author[1]{\mbox{Pushmeet Kohli}}
\author[1]{\mbox{James Kirkpatrick}}
\affil[*]{Equal contribution}
\affil[1]{DeepMind, London, UK}
\affil[2]{Department of Materials Science and Engineering, Massachusetts Institute of Technology, Cambridge, Massachusetts, USA}
% \affil[2]{Google Brain, Mountain View, USA.}

\renewcommand\Authands{ and }

\newcommand{\energy}{e}
\newcommand{\embedding}{v}
\newcommand{\formula}{u}

\date{May 2019}

\maketitle
\\

\begin{abstract}
Machine Learning (ML) has the potential to accelerate discovery of new materials and shed light on useful properties of existing materials. A key difficulty when applying ML in Materials Science is that experimental datasets of material properties tend to be small. In this work we show how material descriptors can be learned from the structures present in large scale datasets of material simulations; and how these descriptors can be used to improve the prediction of an experimental property, the energy of formation of a solid. The material descriptors are learned by training a Graph Neural Network to regress simulated formation energies from a material's atomistic structure. Using these learned features for experimental property predictions outperforms existing methods that are based solely on chemical composition. Moreover, we find that the advantage of our approach increases as the generalization requirements of the task are made more stringent, for example when limiting the amount of training data or when generalizing to unseen chemical spaces.
\end{abstract}

\section{Introduction}

Deep learning has established itself as the dominant method for image recognition\cite{szegedy2017inception}, speech recognition\cite{oord2016pixel}, machine translation\cite{vaswani2017attention} and planning\cite{silver2018general}. More recently, deep learning based methods have been improving the state of the art in physical models of the natural world\cite{carleo2019review}, for example in protein folding\cite{senior2020improved}. The key to deep learning's success in all these domains is that the composition of several layers of a neural network extract highly expressive and abstract features from  data. The downside of such learning approaches is that plentiful data or accurate simulators are required to achieve the impressive milestones cited above. Applying such methods to accelerate materials discovery\cite{schmidt2019review} is limited by the fact that high quality experimental data requires expensive laboratory measurements, meaning datasets usually only cover limited material diversity\cite{butler2018machine}. An alternative source of data is quantum mechanical simulation (e.g. density functional theory (DFT)), which has already been used to create datasets of material properties covering a significantly larger material space. However, DFT methods have large systematic errors\cite{kirklin2015open, Kim2017ExperimentalEnthalpies} due to the limitations of the underlying density functional approximations\cite{cohen2008insights,cohen2012challenges} as well as other effects like the \SI{0}{K} approximation\cite{lany2008semiconductor}. The challenge is therefore to leverage the information in large simulation databases to learn a transferable embedding that enables accurate experimental property prediction even from small amounts of experimental data.

\tx{Considerable advances have been made in demonstrating that neural networks and other machine learning models can regress properties computed from DFT, both in the context of organic molecular systems\cite{duvenaud2015convolutional,faber2017prediction, schutt2018schnet,yang2019analyzing} and for bulk materials\cite{ghiringhelli2015big,isayev2017universal,ouyang2018sisso,Jha2018ElemNet, xie2018crystal, chen2019graph}}. The accuracy of such networks depends crucially on the choice of architecture and on the representation of the inputs. In the case of bulk materials, some models, like the ElemNet\cite{Jha2018ElemNet}, represent materials by their chemical composition alone. However they would be unable to distinguish different phases of materials, where the atomistic structure is different but the chemical composition is the same. Alternatively, approaches based on graph networks\cite{scarselli2009graphnetwork} are capable of using the material's atomistic structure as the input representation. Since the unit cell structure of a material underpins its quantum mechanical properties, the power of such structured models is that they instill physically plausible inductive biases in the network\cite{battaglia2018relational}, which explains why these models can achieve high accuracy\cite{xie2018crystal, chen2019graph}. Physically motivated inductive biases also allow networks to be trained more effectively with small amounts of data.

%\tx{Note for the above revision: I think the point of ElemNet is to prove only using composition can also achieve good performance. And we shouldn't criticize them too much as we are likely to be our reviewers.}

In this work, we develop an atomistic graph neural network that learns a transferrable embedding of materials from DFT calculations to improve predictions on experimental properties. Recently, Jha \emph{et al.}\cite{Jha2019NatureComm} showed such improvements by pre-training ElemNet on large simulation datasets\cite{Jha2019NatureComm}, which only learns from the composition of materials. By learning from the atomistic structures that underpin these simulated values, our approach can more accurately predict experimental quantities, even when trained on small amounts of data. Specifically, we implement these ideas in a practical algorithm to regress experimental energies of formation (EOF)\footnote{Experiments typically measure finite temperature \emph{enthalpies} while simulation focus on zero temperature \emph{energies}; for simplicity we refer to both as energies in this work}. We demonstrate that our network architecture outperforms models that do not use structural information, and furthermore show that it is more data efficient, meaning that the test error degrades in a less pronounced way when limiting the training set or splitting training and testing data in more challenging ways. 
A major difficulty with learning from atomistic structure is that structural information is often not recorded in experimental datasets due to the additional experimental cost of collecting accurate material structures. To overcome this we learn an embedding of atomistic structures into a continuous vector space using simulated structure-property pair data. The embedding for any unknown experimental structure can be obtained by interpolation in embedding space of known nearby structures using an interpolation scheme inspired by the convex hull decomposition in compositional phase diagrams.
Specifically, our network is composed of a graph network 'trunk', which computes an embedding of a material structure, and two 'heads': one to regress experimental values and another for simulation values. Knowledge is transferred from the simulation data to the experiment head by the shared embedding trunk: the trunk and the simulation head are trained on simulation data to learn structure-property relations similar to refs.\cite{xie2018crystal,chen2019graph}, and the trunk and the experimental head are then fine-tuned on experimental data. 

 We find that incorporating structural information allow us to achieve a new state of the art mean absolute error (MAE) of $0.059 \pm \SI{0.004}{eV/atom}$ when predicting the formation energy on an experimental dataset of 1963 compounds. 
Importantly, the method still generalizes well in the small training data regime compared to the baseline which uses a simple multilayer perceptron (MLP) as a trunk and does not include structural information. When training on only 157 experimental points our structured approach achieves 17\% lower mean absolute error (MAE) than the previous state of the art method \cite{Jha2019NatureComm}. We also find improved generalization ability to new chemical spaces, achieving  31\% lower MAE when holding out copper from the experimental training set. Finally, although trained on EOF, we demonstrate that the model provides estimates for the \emph{decomposition} energy (\emph{i.e.} the energy of formation difference with respect to all the other compounds in the chemical space), which are comparable in accuracy to values computed from simulation\cite{bartel2020critical}.

\begin{figure}[h]
\begin{center}
    \includegraphics[width=0.8\textwidth]{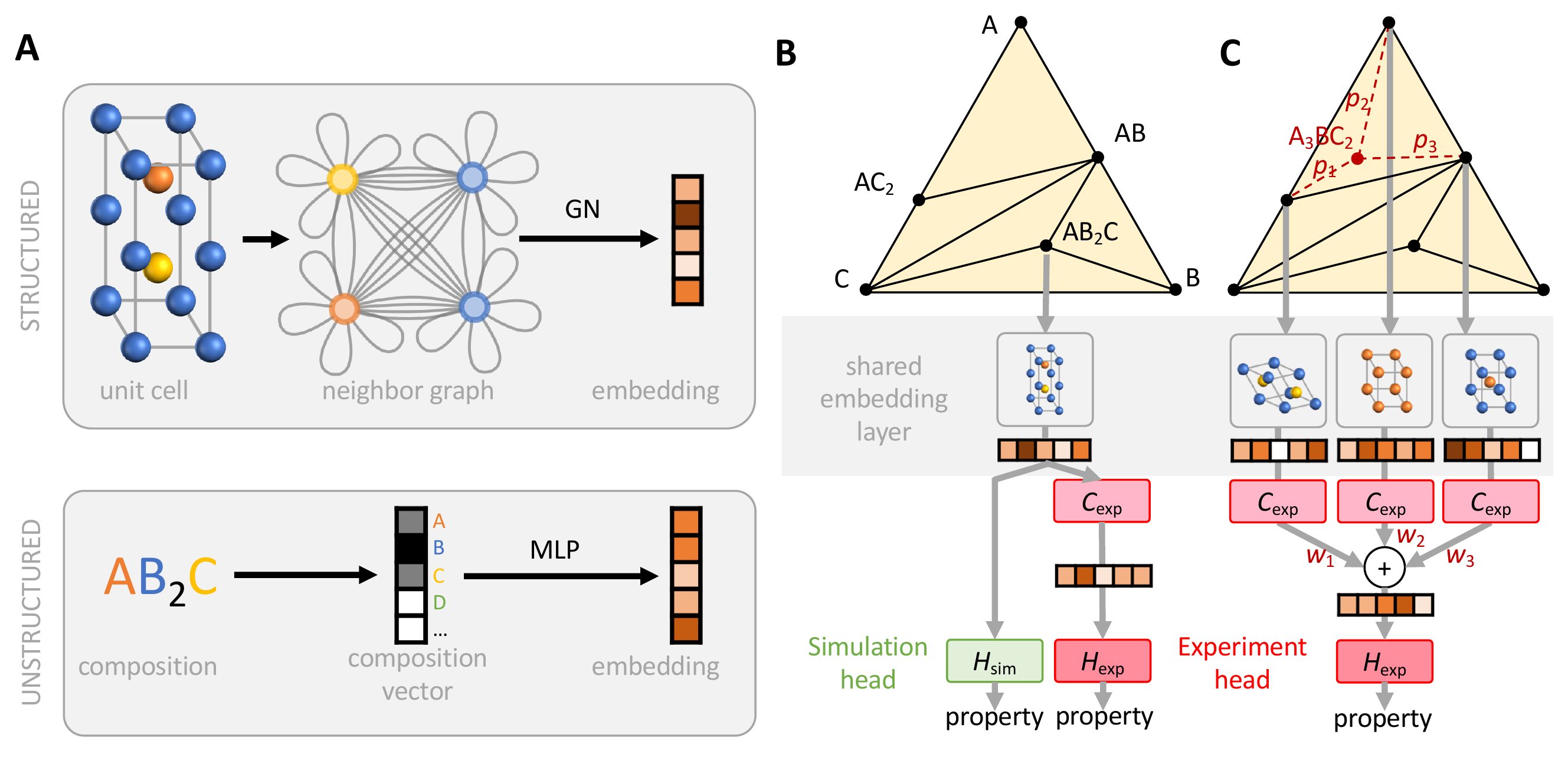}
    \caption{Illustration of the transfer learning process. a) The two families of network compared in this work. Structured approaches (AGN) use a graph representation of a crystal to compute an embedding of the compound. Unstructured approaches (MLP) use a vector representation of a formula to compute an embedding of the compound. b) For compounds in the simulation dataset the simulation head computes a prediction for the simulated property from the embedding. For compounds in the experimental dataset that have a stable correspondence in the simulation dataset, the experimental head computes an experimental energy from the embedding. c) For the experimental compounds that do not have a correspondence in the simulation dataset, a phase diagram is first constructed using the simulated materials (black dots) and their formation energies. The simulated materials are fed into a graph network and a representation for the experiment material is synthesized as a weighted average of the embeddings of its neighbors in the phase diagram. This is used to predict the experiment property.}
    \label{fig:main_illustrative}
\end{center}
\end{figure}

\section*{Results}

\paragraph{Atomistic graph network}

Our aim is to predict experimental formation energies $e_{\textrm{exp}}(x)$ for compounds $x$ in an experimental dataset $\mathcal{E}$ by leveraging the information in another dataset $\mathcal{S}$ containing simulated data. This represents a scenario where researchers hope to experimentally measure the formation energies of all compounds in $\mathcal{E}$, and our goal is to accelerate this process by training a ML model that utilizes a large open simulation dataset $\mathcal{S}$ and a few experimental measurements on some compounds in $\mathcal{E}$.

The simulation dataset $\mathcal{S}$ contains calculations of various physical quantities as well as the physical structure of the material. To leverage both the structure and property information, for every example $y \in \mathcal{S}$, we can construct a graph representation  $g(y)$ of the corresponding crystal by connecting atoms under a certain threshold distance with edges, and adding the atomic number as node features and distances as edge features\cite{xie2018crystal}. This allows us to feed $y$ into a graph network\cite{scarselli2009graphnetwork,battaglia2018relational} $T$ in order to learn an embedding $\embedding_T(y) = T(g(y))$. \tx{Note that a recent work MEGNET \cite{chen2019graph} also uses graph networks to encode material structures}. Supervision for this embedding comes from passing $\embedding_T(y)$ through the simulation head MLP, $H_{\textrm{sim}}$, to produce a predicted energy $\hat{\energy}_{\textrm{sim}}(y)$ and regressing it towards with the simulated energy $\energy_{\textrm{sim}}(y)$.
%Simulated atomization energies are then regressed based on this embedding vector (\autoref{fig:main_illustrative} a and b):
%$$ \hat{E}_{\textrm{sim}}(y) = H_{\textrm{sim}}(e_T(y)) \rightarrow E_{\textrm{sim}}(y). $$
%In this equation, $H_{\textrm{sim}}$ denotes a multi-layer perceptron (MLP) which is specific to the task of predicting simulated properties.
The learned embedding based on the physical structure of material $y$ is the key element that will allow us to transfer from simulation data to experimental data.

When we transfer to the experimental data, we first need to compute an embedding vector $\embedding_T(x)$ for each example $x \in \mathcal{E}$. Note that there is no prior knowledge of the structure of $x$, so we need to find a way to match the learned embedding of structures in $\mathcal{S}$ and $x$. If there is a single \emph{stable} entry $y$ in the simulation database matching the chemical composition of $x$, we \emph{assume} that the physical structure of $y$ is the true structure of $x$, although this assumption may be incorrect as the stability and structure of $y$ are obtained from simulation data alone. In this \emph{stable match} case, we can use the embedding layer $T$ shared from the simulation architecture to compute the embedding for $y$  as an effective embedding for $x$ (\autoref{fig:main_illustrative}b). We pass this through two experiment-specific MLPs: the ``composition'' network $C_{\textrm{exp}}$ and the experimental head $H_{\textrm{exp}}$ to generate the predicted energy as
\begin{equation} 
\bar{\embedding}_T(x) = C_{\textrm{exp}} (\embedding_T(y)), \qquad \hat{\energy}_{\textrm{exp}}(x) = H_{\textrm{exp}}(\bar{\embedding}_T(x)).
%\hat{\energy}_{\textrm{exp}}(x) = \left(H_{\textrm{exp}} \circ C_{\textrm{exp}} \right) (\embedding_T(y)) \rightarrow \energy_{\textrm{exp}}(x), 
\label{eq:gnn_update_exact_match}
\end{equation}
%where $\embedding_T$ is as before and the "composition" network $C_{\textrm{exp}}$ and the experimental head $H_{\textrm{exp}}$ are MLPs.
%In the case where there are several $y$ with this property (typically corresponding to different crystallographic structures), we pick the one with the lowest predicted formation energy $e_{\textrm{sim}}(y)$, corresponding to the stable compound.
If there is no stable compound in the simulation dataset matching the formula of $x$, one way to guess the structure embedding of $x$ is to interpolate the structure embeddings of neighboring compounds in the phase diagram. We decompose $x$ according to the phase diagram construction (obtained with the pymatgen software\cite{pymatgen}). This yields a set of stable materials $y_1, \dots, y_n \in \mathcal{S}^n$ and weights $p_1, \dots, p_n \in (0, 1]^n$ for each compound $x$ (see \autoref{fig:main_illustrative}c). We then compute the effective experimental embedding as:
\begin{align}
\bar{\embedding}_T(x) &= \sum_{i = 1}^{n} w_i C_{\textrm{exp}}(\embedding_T(y_i)), \qquad w_i = \frac{p_i^\alpha}{\sum_{i=1}^{n} p_i^\alpha},
%\qquad {\hat{\energy}}_{\textrm{exp}}(x) = H_{\textrm{exp}}(\bar{\embedding}_T(x)),
\label{eq:gnn_update}
\end{align}
 where $\alpha$ is a parameter controlling how component embeddings are interpolated. Setting it to $0$ averages all stable neighbours uniformly, setting it to $\infty$ uses the stable neighbour with the largest weight only and setting it to $1$ uses the weights from the phase decomposition, unless stated otherwise we use $\alpha=1$. Note that Eq. (\ref{eq:gnn_update}) reduces to Eq. (\ref{eq:gnn_update_exact_match}) in the 'stable match' case, where $n=1$; our ``phase diagram decomposition" procedure is therefore a unified method to produce an embedding from a chemical formula and a structural dataset.
 %and the vector $w$ reduces to $w = (1,)$.
 %The composition network $C_{\textrm{exp}}$ and the decoding head $H_{\textrm{exp}}$ are specific to the training on the experimental datasets, but to achieve knowledge transfer the embedding layer $T$ is shared with the network that predicts the simulation energy.
 Applying the same composition network to each embedding vector $\embedding_T(y_i)$ followed by averaging by the weights $w_i$ respects the fact that the weights $w_i$ and structures $y_i$ form an unordered set and are therefore invariant to permutation of the elements\cite{Zaheer2017_Deep}. We denote the approach described here as \gn{} in the following. We emphasize that this approach is merely an Ansatz which we found to work well in our case, especially as the structure and formation energy of a compound is already often close to the one given by the phase diagram decomposition, allowing our network to learn corrections over a good initial estimate.

In a further refinement to this method (\emph{\gnsimm{}}\footnote{\tx{In the rest of the paper, every model whose name has a + has the simulated energies as an additional input.}}), we feed as an additional input to $C_{\textrm{exp}}$ the simulated energies of each material, \emph{i.e.} the experimental embedding is computed as: $\bar{\embedding}_T(x) = \sum_{i = 1}^{n} w_i C_{\textrm{exp}}([\embedding_T(y_i), \energy_{\textrm{sim}}(y_i)])$ (where $[,]$ denotes the concatenation operation). This provides more information to the network, allowing it to learn corrections over the DFT predictions depending on the structure of the compounds, or of its phase diagram neighbours\cite{zhang2018strategy}.

Both the simulation and the experimental heads are trained at the same time; the total loss that we minimize by gradient descent is
\begin{equation} \label{eq:loss}
    \mathcal{L}(t) = \omega_{\textrm{sim}}(t) \mathbb{E}_{x \in \mathcal{S}} \left[ \left | \hat{\energy}_{\textrm{sim}}(x) - \energy_{\textrm{sim}}(x) \right | \right] + \omega_{\textrm{exp}}(t) \mathbb{E}_{x \in \mathcal{E}} \left[ \left | \hat{\energy}_{\textrm{exp}}(x) - \energy_{\textrm{exp}}(x) \right | \right],
\end{equation}
where $t$ represents the training iteration, and $\omega_{\textrm{exp}, \textrm{sim}}(t)$ control the relative weight of the experimental loss with respect to the simulation loss. Note that different schedules for $\omega_{\textrm{exp}, \textrm{sim}}(t)$ allow us to trial either a multi-task learning and a fine-tuning approach. In practice we found that annealing $\omega_{\textrm{sim}}(t)$ to a small value over a fixed schedule during training works best.

\paragraph{Baselines}
We compare our proposed approach with the following baselines.

\textit{DFT} There have been many efforts to compute the formation energies of inorganic compounds using various forms of density functional theory (DFT) with known structures (similar to the stable match cases) \cite{materialsproject,oqmd,curtarolo2012aflow}. The formation energies are usually obtained by computing the total energy differences between the compound and their elemental reference states, using structures obtained from the Inorganic Crystal Structure Database (ICSD) \cite{hellenbrandt2004inorganic}. Since some reference states like \ce{O_2}, \ce{N_2} are in gas phase, their references energies are adjusted by fitting with experimental formation energies. A study by Kirklin et al. reports MAEs between \SI{0.081}{eV/atom} and \SI{0.136}{eV/atom} for DFT computed formation energies depending on different fitting schemes in the Open Quantum Materials Database (OQMD)\cite{kirklin2015open}. \tx{To evaluate the DFT error in our dataset, we further perform a linear fit from DFT to experiment formation energies to correct the systematic underestimation of DFT calculations \cite{jain2011formation}. We find a MAE of 0.145 eV/atom for a total number of 1499 stable match compounds. There are also 464 compounds that do not have a stable match in our simulation dataset. We estimate the formation energies of these compounds by averaging the DFT formation energies of their neighbors $\sum_{i = 1}^{n} w_i \energy_{\textrm{sim}}(y_i)$. We find the MAE between DFT and experiment formation energies for the compounds without a stable match to be 0.158 eV/atom, and the MAE for all compounds to be 0.148 eV/atom.}

\textit{MLP without simulation data} The most direct approach is to train an MLP ony on the experimental data, similarly to the ElemNet model\cite{Jha2018ElemNet, Jha2019NatureComm}. The network directly predicts $\hat{\energy}_{\textrm{exp}}(x)$ from its formula. We used a 3-layer MLP, and found that this simpler architecture performed better than the more complex ElemNet model in the low data regime that we are considering in this work.

\tx{\textit{AGN without simulation data} This approach uses our AGN architecture but does not train the simulation head ($\omega_{\textrm{sim}} = 0$ in Eq. (\ref{eq:loss})). The model does not access simulated properties, but it still uses the structures and phase diagram matching information from the simulation dataset.

\textit{Automatminer} A recent approach\cite{dunn2020benchmarking} that automatically selects material descriptors and machine learning algorithms based on the dataset and model performance. The approach is one of the state-of-the-art methods based on human-designed descriptors. The framework is highly customizable and we use the ``express'' setting in our study. The approach do not have access to the simulation data and structure information.}

\textit{\mh{}} A more recent approach\cite{Jha2019NatureComm} combines the ElemNet modelling strength with the abundance of simulation data by performing transfer learning between the simulation dataset and the experimental dataset. For a more direct comparison with our method, we implement this baseline by training two separate heads that come after a common MLP embedding layer, i.e. $\hat{\energy}_{\textrm{exp}}(x) = H_{\textrm{exp}}(M(x))$ and $\hat{\energy}_{\textrm{sim}}(x) = H_{\textrm{sim}}(M(x))$, where $M, H_{\textrm{exp}}$ and $H_{\textrm{sim}}$ are MLPs, and the loss is as in Eq. (\ref{eq:loss}). We also use shallower networks than in the original ElemNet model as we find this slightly improves on the previously published results.

\textit{\mhsim} This approach is similar to the MLP, but the transfer head takes as additional input the average feature $\sum_{i=1}^n w_i \energy_{\textrm{sim}}(y_i)$, where the weights are computed as in Eq. (\ref{eq:gnn_update}). 

% \ag{just remove all of this?}
% It was recently suggested that using an attention based model on formulas could improve results when training from a stochiometric formula\cite{goodall2019roost}, but in our study we did not find that this performed better than the simpler \mh{} approach, so we do not report it here. We believe that the apparent discrepancy with previously reported results\cite{goodall2019roost,bartel2020critical} might be due to the difficulty of training the original ElemNet model. \tx{I tried to soften this sentence a little since it was criticized by reviewer 3. Although I disagree with the reviewer, we might face a similar (or the same) reviewer in our resubmission.}%Moreover, we hypothesize that a sufficiently well tuned MLP should be able to achieve similar performance to a graph based approach which only gets a graph representation of a formula as input, as representing the latter as a graph does not provide additional information to the model. \victor{Probably a bit too long}
% \ag{$\backslash \rm remove$}

\paragraph{Datasets}
We use the same experimental dataset as in the most recent state of the art work\cite{Jha2019NatureComm}, the SGTE Solid SUBstance dataset\cite{sgtebook}. This dataset initially contains 2090 compounds. Performing the same standard filtering as was done in prior work\cite{Jha2019NatureComm} (i.e. removing compounds with formation energy more than 5 standard deviations away from the mean), we obtain a dataset of 1963 compounds. We note that the resulting dataset contains an important fraction of duplicate compounds: the dataset only contains 1642 unique formula, with 197 compounds duplicated at least three times. For the simulation dataset, we use data retrieved from the Materials Project\cite{materialsproject}, with no filtering -- giving us a dataset of 120,612 compounds.
%Contrarily to what was previously reported\cite{Jha2019NatureComm}, we do not observe a strong dependency of the results on the simulation dataset used (see \supplementary~for details).

\paragraph{Performance on the full dataset} 

% \begin{figure}[h]
% \begin{center}
%     \includegraphics[width=\textwidth]{main_results_select.pdf}
%     \caption{(a) Mean absolute error (MAE) on the experimental test set. The error bars are computed as the standard deviation over the 10 different validation splits, for the 10 different models that we trained. A tentative estimate of the experimental error (a lower bound on the achievable error) can be obtained using the duplicates of the dataset, yielding a value of 0.009 eV/atom for the RMSE (similar MAE), twice lower than the reported value of 0.0225 ev/atom reported for another, similar experimental dataset\cite{Kim2017ExperimentalEnthalpies}. \ag{Remove Fig2b} (b) a histogram of errors for the different methods presented in the text. We use the MAE on the validation splits to increase the amount of data, but the picture is similar on the test MAE.}
%     \label{fig:main_results}
% \end{center}
% \end{figure}

\begin{table}[]
    \centering
    \caption{Mean absolute error (MAE) on the experimental test set. The error bars are computed as the standard deviation over the 10 different validation splits, for the 10 different models that we trained.}
    \begin{tabular}{p{2.5cm}|c}
        Method & MAE (eV/atom)\\
        \hline
        DFT (match) & 0.145 \\
        DFT (all) & 0.148\\
        MLP w/o sim & $0.128\pm{0.004}$\\
        AGN w/o sim & $0.128 \pm{0.003}$\\
        Automatminer & $0.128 \pm 0.007$ \\
        MLP & $0.065 \pm 0.003$\\
        MLP+ & $0.064 \pm 0.002$\\
        AGN & $0.063 \pm 0.003$\\
        AGN+ & $\bf 0.059 \pm 0.004$\\
    \end{tabular}
    \label{tab:mae-exp}
\end{table}

We first compare the performance of our approaches on the full experimental dataset. We create a train test split of our dataset in proportions 8:2, and then further redivide the training set in 10 splits, one of which is used as validation during training and for early stopping. Because of the presence of repeated compounds, we randomly split the compounds in a way that places compounds with the same formula in the same split of the dataset. This avoids the issue of contamination of the test set by the training set.

We report performances of our models and several baselines in \autoref{tab:mae-exp}. The MLP without simulated data baseline reaches a MAE of 0.13 eV/atom, similar to the previously reported values in Ref.\cite{Jha2019NatureComm}. \tx{Automatminer performs similarly in the dataset because it also does not have access to the simulation data, and it seems that the human-designed features do not improve the performance of the model compared with a simpler MLP.} All the approaches that perform transfer learning obtain a significantly lower error, compared with both DFT, the MLP without simulated data, and automatminer. This shows that transferring information from simulation can improve the prediction on experimental properties, consistent with observations before\cite{Jha2019NatureComm}. Interestingly, the errors of these models are significantly smaller than the MAE between DFT and experiments even after a linear fit for all stable match cases (0.145 eV/atom), indicating that they might learn non-trivial corrections to the DFT formation energies, in agreement with previous studies\cite{zhang2018strategy}. We find that the graph based approaches (\gn{} and \gnsimm{}) outperform the unstructured approaches, yielding a new state of the art mean absolute error of $0.059 \pm 0.004$ eV/atom, compared with $0.07$ eV/atom in Ref.\cite{Jha2019NatureComm}. We also observe that giving the transfer model access to the values computed by DFT systematically improves results, both for the MLP and for the graph network based approaches. \tx{Directly feeding the DFT computed energies of formation to the networks without performing transfer learning yields performance which are close but slightly worse than our transfer approaches (including those that don't have access to DFT data), as demonstrated in Supplementary \autoref{fig:app_no_transfer}. This emphasizes the role of \emph{feature learning} which happens during the transfer learning: the networks can learn, for each material, a representation from the simulations which is richer than the scalar number it is regressing to.}

% \ag{Remove fig 2b?}
% In \autoref{fig:main_results}b, we show the histogram of errors for the different methods, demonstrating that all the transfer based method have similar tails but that the ones that have access to DFT data (\mhsim, \gnsimm) have their modes shifted to lower values \ag{I find this shift very unconvincing in the figure}. On this plot we also compare directly with the prediction that would naively be obtained from DFT, $e = \sum_{i=1}^n w_i e_{\textrm{sim}}(y_i)$ (with the notations of Eq. (\ref{eq:gnn_update})), highlighting that the error of this method is significantly larger than the one of the learned models \ag{we already knew that the error for DFT was higher from fig2a}. To reconcile this fact with the improved performance of the methods when augmented with the ground truth from DFT, we hypothetize that these methods are leveraging the systematic character DFT errors, and are learning corrections over the DFT predictions, in agreement with previous studies\cite{zhang2018strategy}. \ag{we already said that we're learning corrections}
% \ag{$\backslash \rm remove$}

\paragraph{Generalization performance}

\begin{figure}[h]
\begin{center}
    \includegraphics[width=\textwidth]{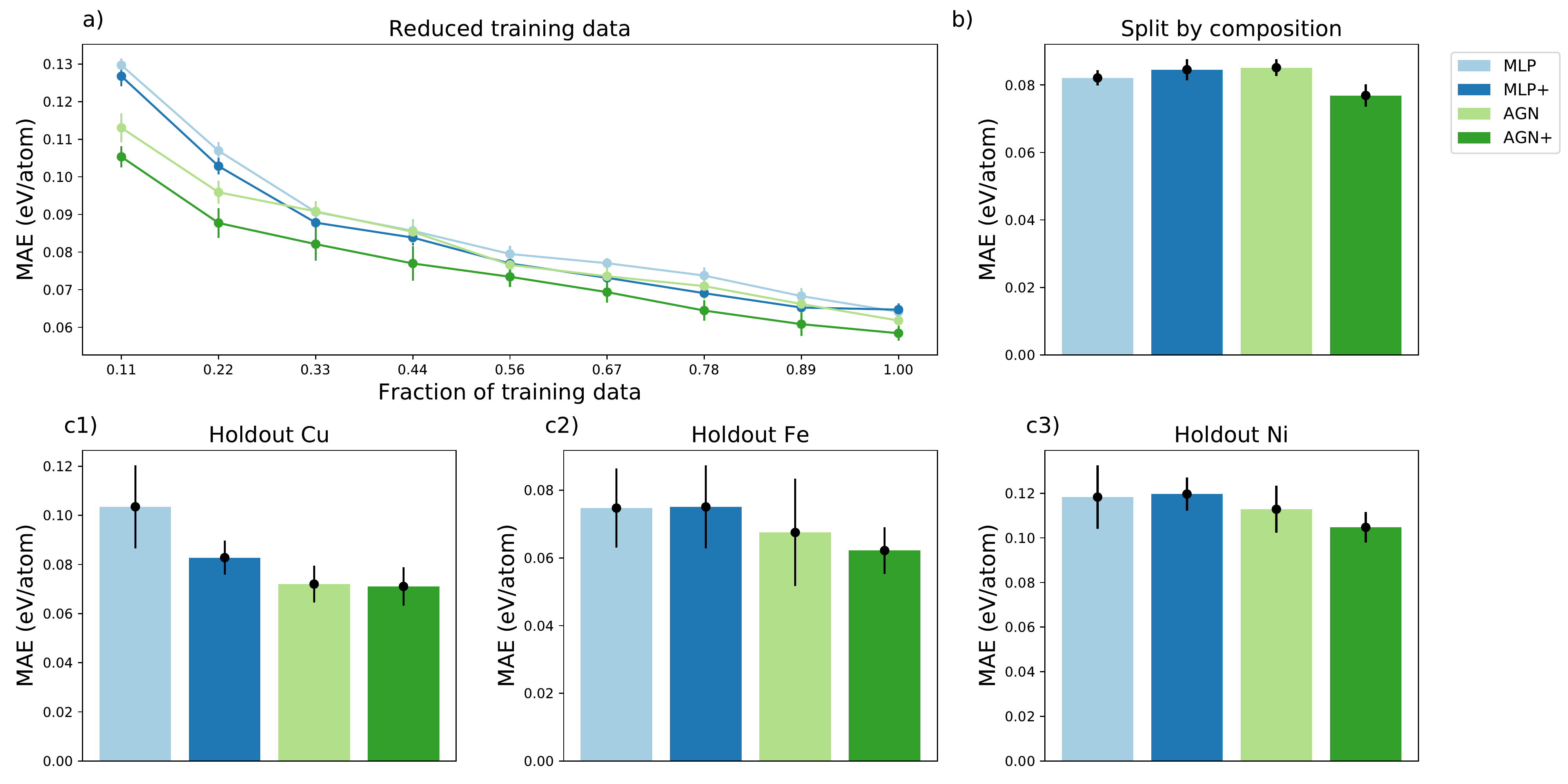}
    \caption{Generalization performance on the experimental dataset. (a) MAE as a function of the fraction of experimental training data used (the size of the validation set is kept constant and is not counted as training data). The performance of the MLP without simulation data is not shown here as it deteriorates even quicker than the transfer based approach. (b) MAE on a split of the dataset that respects composition groups (see text for details). (c) performance on splits of the experimental datasets for which all compounds in the test set contain a chemical element not present in any of the training data (from left to right, copper, iron and nickel). The MLP without simulation data approach is not shown here as it yields results with an error several times larger.}
    \label{fig:low_data}
\end{center}
\end{figure}

The key issue we wish to address is how \emph{data efficient} our model is as we decrease the size of the experimental training set while keeping the size of the simulation datasets unchanged, and how well it generalizes beyond the training set. Data efficiency is vital in material science applications where datasets of experimental properties have few examples and collecting thousands of datapoints might be unfeasible for all properties of interest. Our hypothesis is that the strong inductive bias of our structured approaches should require less data to achieve the same error on a fixed test set. In  \autoref{fig:low_data}a, we show the performance of the various models with reduced experimental training set size, and we find that the performance of the structured approaches degrades more gracefully as the amount of training data is reduced. For instance, \gnsimm{} achieves the same error with only 157 examples as \mhsim{} does with twice the data, and when using less than 400 training examples the performance of the \gn{} method (without direct access to simulated DFT energies) outperforms the \mhsim, which has access to DFT. In addition, even with just 157 training examples, the MAE of the \gnsimm{} method is still significantly smaller than the MAE of DFT with respect to the experiments in both stable match and all cases. This shows that structured approaches can correct DFT computed formation energies even with limited data from experiments.

Beyond data efficiency, an important property required to use machine learned models for materials discovery is the ability to generalize to test sets that contain examples that are significantly different from the training set. In \autoref{fig:low_data}b, we consider a train/test split where the test set does not contain materials involving the same set of atoms as in the training set (as recently proposed\cite{bartel2020critical}). For instance, if $\textrm{Hf}_2\textrm{Si}$ is in the test set, then no other materials composed of both hafnium and silicon (such as $\textrm{Hf}_3\textrm{Si}_2$ and $\textrm{HfSi}$) would be allowed in the training set. We observe that the corresponding task is harder for all methods, but that the graph based approach with access to the simulation data performs best in this setup.
An even more extreme test of generalization is when all materials involving a certain element are removed from the experimental training set. In the bottom row of \autoref{fig:low_data} we show the test error on compounds containing a particular element when the network has never seen an experimental compound containing that element during training. We picked copper, iron, and nickel as the elements for this generalization test, and find that the gap in performance between the structured approaches and the unstructured ones is wide: up to 30\% in the case of copper. Note that these generalizations are possible because the models are learning features from the simulation datasets that has a broader range of materials; these features are useful for experimental property prediction even on elements that are unseen in the experimental dataset.

The fact that \gn{} and \gnsimm{} outperform \mh{}  and \mhsim{}  in both data efficiency and generalization shows that the structured approaches are learning more powerful embeddings from the DFT data. It also indicates that such approaches are likely to be useful in the materials discovery scenario, where we might need to extrapolate accurately from few experimental points even when materials are significantly different from the ones that have already been measured.

\paragraph{Analysing the choice of interpolation method.}

\begin{figure}[h]
\begin{center}
    \includegraphics[width=\textwidth]{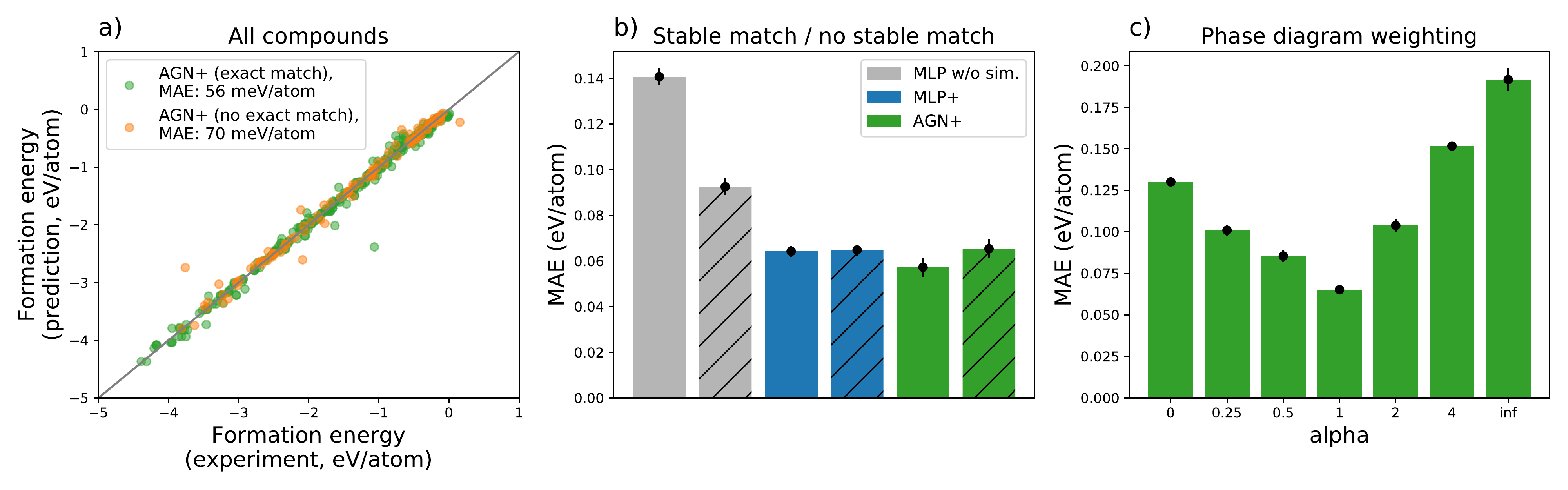}
    \caption{(a) predicted formation energy for the \gnsimm{} method, for the compounds with a stable match (green) and with no stable match (orange), plotted against the experimental formation energy. (b) MAE on the test dataset, split by formula, depending on whether an experimental compound has a stable match in the simulation dataset (solid bars) or has none (dashed bars). For reference, the mean (resp. median) DFT error in the single neighbor case is 0.47 eV/atom (resp 0.14 eV/atom). (c) MAE as a function of the parameter $\alpha$ of Eq. (\ref{eq:gnn_update}). In both panels, each bar shows the average mean absolute error for 10 models trained with 10 different heldout validation splits. Error bars are computed as the standard deviations over the 10 different validation splits, for the 10 different models that we trained. }
    \label{fig:results_analysis_1}
\end{center}
\end{figure}

In the absence of exact experimental structures we rely on interpolation to produce effective embeddings from neighbours.
%In order to compute an embedding vector for the materials in the experimental database, we need to need to associate each chemical formula with a set of weights and simulated structures, because the true structures of experimental materials are not known. Is our choice of using phase diagrams decomposition reasonable? 
To investigate the efficacy of these interpolated embeddings and the effect of $\alpha$ in \autoref{eq:gnn_update}, we study the mean error for materials with or without an stable match in the simulation database (\autoref{fig:results_analysis_1}a and b). The MLP without simulation data baseline (which is indifferent to whether the experimental data has or has not a stable match in simulation data) shows a better performance on the data without a stable match indicating that this subset of our test set is intrinsically `easier' (see also Supplementary Figure \ref{fig:results_analysis_app}). Conversely, the graph based approaches find it slightly harder because without a stable match we rely on an assumption that phase diagram neighbours give an effective embedding that is a good approximation to the true structure embedding.
%simulation data is intrinsically different: in the absence of a stable match the graph net operates on graph \emph{potentially} exactly describing a crystallographic structure of the compound, while in the former case we have to rely on the phase diagram approach of Eq. (\ref{eq:gnn_update}) with a vector $w$ of length strictly more than one.
%We find that the performance of the graph network correspondingly degrades compared to the stable match case.
We present a more detailed analysis in the \supplementary{} where we show that the error increases as soon as there is not stable match, but does not strongly vary with the number of neighbors in the phase diagram afterwards.
%The central panel of \autoref{fig:results_analysis_1} presents a more detailed analysis of this finding. We plot the error against the inverse participation ratio $ipr(w) = \left(\sum_{i=1}^n w_i^2 \right)^{-1}$ -- a continuous measure of the number of neighbors in the phase diagram (for instance, if a compound is uniformly spread over $n$ neighbors, $w=(1/n, 1/n, \dots, 1/n)$, then $ipr(w)=n$.). This provides a finer analysis of the error in the case where there is no exact match in the simulation  We see that error is relatively independent of the inverse participation ratio, with the main increase happening very near $ipr(w)=1$. This shows that the predictive hardness increases more sharply when the crystal structure becomes unknown, but that beyond that, the number of compounds in the phase diagram and the distance to the nearest compound do not have a strong influence.

Given that we find stable matches easier, it is reasonable to ask if the phase diagram decomposition provides any useful information at all. To investigate this, in \autoref{fig:results_analysis_1}c we show the effect of varying the parameter $\alpha$ used to compute the weights $w$ in Eq. (\ref{eq:gnn_update}). This allows to interpolate between the limit $\alpha=0$, which corresponds to a uniform mixture of the phase diagram neighbor embeddings, and $\alpha \to \infty$, which corresponds to only considering the nearest neighbor(s) embedding(s). We find that the mean absolute error is lowest when $\alpha=1$, which corresponds to a picture where weighted averages of properties using weights from the phase diagram decomposition provide a good proxy for the property of a compound. This is consistent with the observation that the difference between the energy of a compound and the energy obtained by averaging its phase diagram neighbors (the energy of decomposition) is typically one order of magnitude smaller than the EOF, so the phase diagram average should be a good proxy for the EOF.

\paragraph{Energy of decomposition}

\begin{figure}[h]
\begin{center}
    \includegraphics[width=\textwidth]{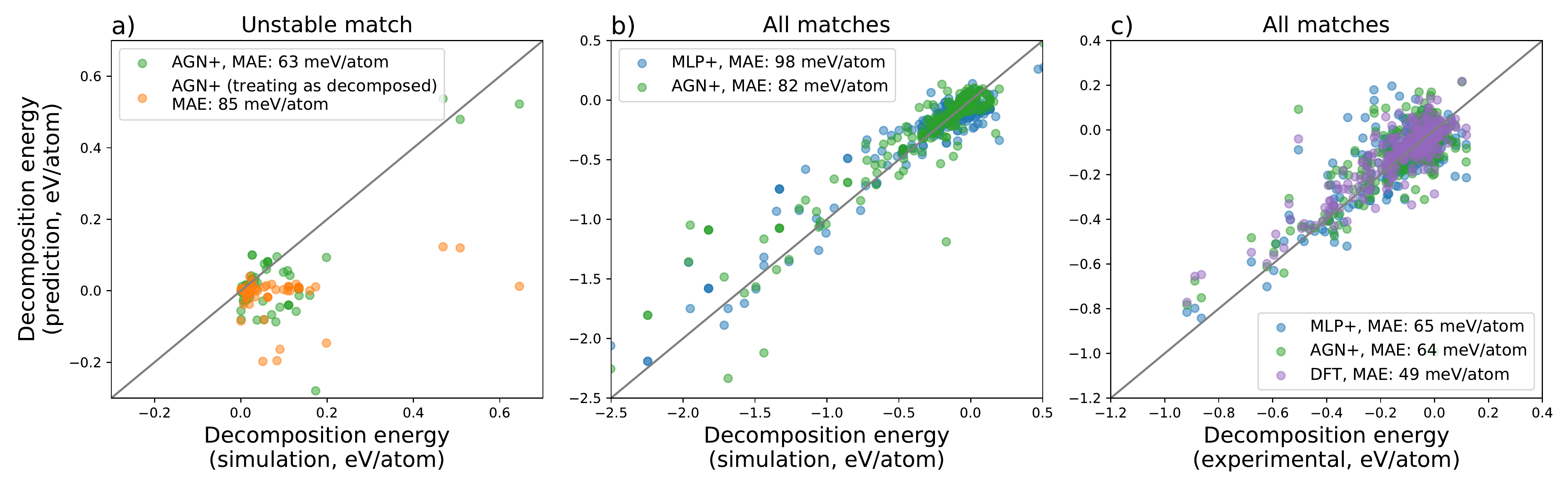}
     \caption{(a) for the matches in the simulation dataset which are unstable, comparison of the output of the pre-trained network obtained from the phase diagram decomposition with the one obtained by feeding the structure of the unstable compound (note that the network was trained using the former procedure). (b) predicted decomposition energy for all the compounds in the test set that have a match in the simulation dataset (including the unstable ones), against the values predicted by simulation. The pre-trained graph network is fed the crystal structure of the unstable compound where applicable. (c) predicted decomposition energy (pre-trained network and DFT) against experimentally determined ones for the experimental compounds that are part of a non-trivial phase diagram in the intersection of the experimental and of the simulation data. In all cases, the network had been trained with our main procedure on the formation energy prediction task.}
    \label{fig:results_analysis_2}
\end{center}
\end{figure}

While available experimental datasets typically contain the energy of formation, the relevant quantity for the stability of a material is the energy of \emph{decomposition}, i.e. the energy of formation with respect to all the other compounds, not just isolated constituents. A negative value indicates that a material is stable, while a positive value indicate it is unstable or metastable. This is typically one order of magnitude smaller and is often not correctly predicted by machine learning models. With the exception of graph based approaches, those models tend to fail to capture the subtle correlations in formation energies required for a good decomposition energy prediction, while DFT provides a more competitive baseline because of cancellation of its systematic errors\cite{bartel2020critical}. The error of our method on the experimental energy of formation is comparable to the error of models that give predictive approximations to the \emph{simulated} energy of decomposition; we therefore ask whether our structured models have sufficient correlations in their errors to potentially be a good predictor of the \emph{experimental} energy of decomposition. \tx{While the true experimental energy of decomposition is often unknown, 
we perform two indirect evaluations to estimate the performance of our proposed method.}

The first indirect evaluation is to compare the predicted experimental energy of decomposition (see Methods) to the one computed from DFT. On \autoref{fig:results_analysis_2}a, we investigate whether the phase diagram decomposition method is capable of obtaining the fine-grained estimate of the energy of formation that would lead to a good energy of \emph{decomposition} prediction, by looking only at the compounds whose formula matches an unstable compound in the simulation dataset\footnote{Since they exist in the experiment dataset, they are either metastable or correspond to a stable phase that is not in the simulation dataset}. For those, we have two options: either feed the graph network the phase diagram decomposition of the compound (following Eq. (\ref{eq:gnn_update})), or feed it the crystal structure of the most stable of its unstable counterparts (following Eq. (\ref{eq:gnn_update_exact_match})). \tx{We find that the predicted decomposition energy correlates with the simulated one only in the latter case, although the correlation is not very strong due to the narrow energy distribution of most compounds.} We see this as an evidence of the correlation between the network errors on neighboring compounds when their structures are known. Interestingly, results on the energy of formation do not improve when following this same procedure, in agreement with the fact that the energy of the hull provides a good approximation to the energy of formation. 

In \autoref{fig:results_analysis_2}b, we report the predicted experimental energy of decomposition against one computed from DFT for \emph{all} the compounds that have a match in the simulation dataset (keeping the procedure of the previous paragraph for the unstable ones). We find that in this test  the graph based approach achieves a lower error than the unstructured approaches; the median error of the \gnsimm{} model (0.037 eV/atom) is under the typical DFT error (reported\cite{bartel2019} around 0.07 eV/atom on 646 reactions from the Materials Project data) and therefore it is impossible to decide whether our model has learned a predictor of the decomposition energy which is slightly worse, equivalent or strictly better than the estimate that can be obtained from DFT.

\tx{We consider the second indirect evaluation in \autoref{fig:results_analysis_2}c by comparing the predicted experimental energy of decomposition against the ones computed from experimental formation energies. This time we compute phase diagram decompositions only from the compounds present in both the simulation and the experimental dataset (we use the test set and the 10 heldout validation sets to maximize the number of generated phase diagrams).} Due to the sparsity of the experimental data, the generated phase diagrams may be incorrect, but the values attached to them are experimentally valid. We find that the DFT computed values provide the best estimate for this metric, but that network based approaches provide estimates with an error of the same order of magnitude (both the \mhsim{} and \gnsimm{} have the same error for this metric).

\section*{Discussion}

We have demonstrated that predictions on experimental material datasets can be systematically improved by transferring both structural information and DFT computed properties from a simulation dataset. 
By comparing with several baselines we find that including structural information improves both the prediction performance and generalization ability of the method, consistent with our initial hypothesis that graph networks have an appropriate inductive bias for the computation of material properties. In particular, we find that using a structured approach is helpful in hard generalization tasks, for example when the training set size is limited or when the test set is very distinct from the training set. 
%It is worth noting that the way that we include the structural information of a material is simply an approximation. 
We have found that the phase diagram decomposition ($\alpha=1$) is the optimum approach for predicting the energy of formation, but it is hard to determine whether this would remain true for all properties. We think the reason why the approach works well empirically might be because the simulation dataset covers a significant part of materials space and most materials are more likely to be decomposed to structurally similar compounds. In addition, it is known that the ground state structure determined by the DFT may not be the true structure of a compound, yet the systematic improvements in performance suggests that even imperfect structural information might still be helpful in predicting experimental properties. \tx{This highlights the potential to augment experimental datasets with DFT computed structures using crystal structure search algorithms\cite{pickard2011ab,wang2010crystal,lonie2011xtalopt}}. It also underlies the hypothesis that structure determination is a foundational subject in theoretical material science and that advances in this field would have repercussions also for property prediction based on learned neural network methods. 

\tx{This approach is a first attempt to incorporate material structure from atomistic simulations as a physics motivated inductive bias to improve predictions on small experimental datasets, but there remains many possibilities for further improvements. First is to extend to predicting more complicated material properties like band gap, elasticity, and thermal conductivity, especially for those compounds without a stable match of simulation structures. We have shown that interpolating the structural representations according to the phase diagram weights is a good inductive bias for predicting formation energies, but better methods might be proposed for other properties with very different characteristics. Another avenue for improvement is to achieve transfer learning between different properties, since there are many experimentally measured properties that are hard to simulate quickly. There are already several studies aiming to transfer between related material properties \cite{yamada2019predicting,sanyal2018mt,chen2019graph}, but transferring to a highly different property remains challenging and may require other physics motivated inductive biases.}

\section*{Methods}

\paragraph{Input preparation}

The unstructured models take the formula $\formula{}(x) \in \mathbb{R}^{100}$ of a compound $x$ as input (where $x$ can belong either to the simulation or to the experimental dataset), obtained from the stochiometric composition of $x$ and normalized such that $\sum_{i=1}^{100} u_i(x)=1$ (none of our compounds involve an atom with atomic number greater than 100).

The graph representation $g(y)$ for compounds $y \in \mathcal{S}$ is created as follows\cite{xie2018crystal}:
\begin{itemize}
    \item Each node corresponds to an atom in the crystal unit cell, and has the corresponding atomic number as a one hot feature in a vector of dimension 100,
    \item Edges are added between pairs of nodes at a distance less than $\delta$, with at most $n$ edges outgoing each node (we used $\delta=\SI{7}{\angstrom}$ and $n=12$),
    \item Each edge has the distance between the two nodes that it connects as a scalar feature
    \item The graph has the chemical formula of the compound, $\formula{}(x)$, as an additional "global" feature.
\end{itemize}
The targets of the networks are normalized to have zero mean and variance one based on the simulation data, i.e. the simulation and experimental head are MLPs followed by a rescaling $x \to (x - \alpha) / \beta x$, where $\alpha = \mathbb{E}_{x \in \mathcal{S}} \left[ e_{\textrm{sim}}(x) \right]$ and $\beta^2 = {\rm Var}_{x \in \mathcal{S}} \left[ e_{\textrm{sim}}(x) \right] $

\paragraph{Models architecture} We optimized our hyper-parameters on the MAE on the validation set for the random split of the dataset. For the MLP without simulation data, we used a MLP with 3 layers of 128 neurons. For the MLP approaches, we used an embedding layer $M$ with 3 layers of 128 neurons; the heads have a single hidden layer of 16 neurons. For the AGN approaches, we use a graph network\cite{battaglia2018relational} with node, edge and global models each made of 2 layers MLP with depth 64. The composition model $C_{\rm exp}$ is a linear layer with dimension 32 followed by a ReLU. The experimental head $H_{\textrm{exp}}$ consists of an MLP with 1 hidden layer of depth 16 and the simulation head is a simple linear network. We used Rectified Linear Units (ReLU) activations and layer normalization\cite{ba2016layer} in all our models.

\paragraph{Training and evaluation} We train all our methods with an Adam optimizer\cite{kingma2014adam} and a learning rate which is decayed from $2\times10^{-3}$ to $2\times10^{-4}$ over a fixed schedule of $N$ iterations, where $N$ was adjusted to the method: we use $N=2\times10^6$ for the \gn{} model, $N=10^7$ for the \gnsimm{} model, and $N=2\times10^5$ for the \mh{} and \mhsim{} models. We use a batch size of 16 for the MLP without simulation data and for the simulation head of the MLP and AGN approaches, and a batch size of 2 for the experimental head of the MLP and AGN approaches. The schedule over the loss, $\omega_{\textrm{sim}}(t)$ in Eq. (\ref{eq:loss}), is a step function with value $\beta$ ($\beta^2 = {\rm Var}_{x \in \mathcal{S}} \left[ e_{\textrm{sim}}(x) \right]$) for $t \leq 2\times 10^5$ and $0.1 \beta$ for $t > 2\times10^5$, while we used a constant $\omega_{\textrm{exp}}(t) = 0.1 \beta$.

\paragraph{Energy of decomposition} We compute the energy of decomposition on pre-trained networks. We obtain this quantity by computing the phase diagram of a given compound $x$ in the simulation dataset (for \autoref{fig:results_analysis_2}a,b) or experimental dataset (\autoref{fig:results_analysis_2}c) from which we have removed $x$. We feed the vertices $y_1, \dots, y_n$ of the phase diagram through our trained networks and compute the energy hull of the phase diagram from it. The decomposition energy is obtained by comparing this hull energy with the formation energy of $x$.

\section*{Data availability}
No data sets were generated during current study. All the data sets used in the current study are available from their corresponding public repositories-- Materials Project (\url{https://materialsproject.org}), and experimental observations (\url{https://github.com/wolverton-research-group/qmpy/blob/master/qmpy/data/thermodata/ssub.dat}).
\section*{Code availability}

 We will make the graph based model and the code necessary to create the graph inputs publicly available upon publication.
 
 \section*{Acknowledgements}
 
 We would like to thank Trevor Back, Tim Green and Alvaro Sanchez-Gonzalez for useful discussions related to this work.
\subsection{Bibliography}
\putbib[main]
\end{bibunit}

\newpage
\setcounter{page}{1}
\begin{bibunit}[naturemag]
\beginsupplement
\section*{Supplementary material}

% \paragraph{Performance on dataset with duplicates}

% In \autoref{fig:app_duplicates} we report the performance of our models on the dataset originally used in prior work\cite{Jha2019NatureComm}, where duplicates in the experimental test set can appear both in the training and test sets. We observe that the performance of the MLP model improves more in that setting, and reaches a performance under the previously reported results.
% \victor{TODO}
% \begin{figure}[h]
% \begin{center}
%     \includegraphics[width=0.8 \textwidth]{main_results_select_a.pdf}
%     \caption{Mean absolute error (MAE) on the experimental test set. The error bars are computed as the standard deviation over the 10 different validation splits, for the 10 different models that we trained. The solid horizontal line represents the previous state of the art result\cite{Jha2019NatureComm} and the dashed line the value reported in the same work for the MLP without simulation data baseline.}
%     \label{fig:app_duplicates}
% \end{center}
% \end{figure}

\paragraph{Performance on simulation data}

\begin{figure}[h]
\begin{center}
    \includegraphics[width=0.6 \textwidth]{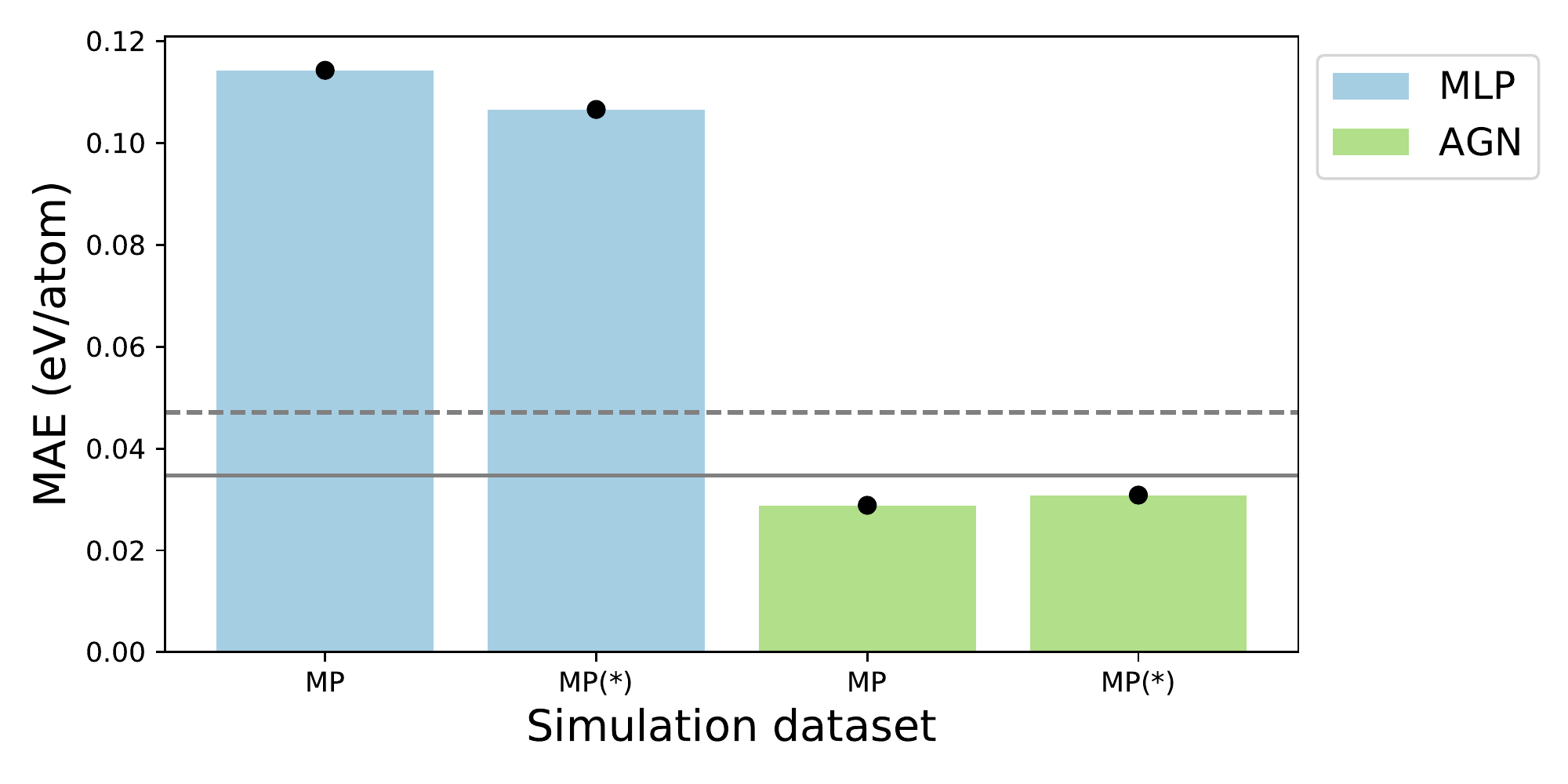}
    \caption{Mean absolute error on a simulation test set as a function of the training dataset used: full Materials Project (MP) or the dataset where compounds with the same formula are removed (keeping only the most stable one) on simulation data (MP*). The error bars are computed over 10 different validation splits, for 10 different models that we trained.}
    \label{fig:app_simulation}
\end{center}
\end{figure}

In \autoref{fig:app_simulation} we report the performance of our models when trained purely on simulation data. Our errors are comparable to those previously reported\cite{xie2018crystal,bartel2020critical} for the structured approaches and worse for the unstructured approaches -- presumably because of the simpler network architecture that we used to optimize transfer learning results. As expected, for the unstructured approaches the performance is better when we remove compounds with different crystallographic structure but same stochiometric formulas (as the MLP approach cannot distinguish between those), while the structured approaches are less effected by this curation of the datasets.

\paragraph{Performance of non-transfer approaches}

\begin{figure}[h]
\begin{center}
    \includegraphics[width=0.6\textwidth]{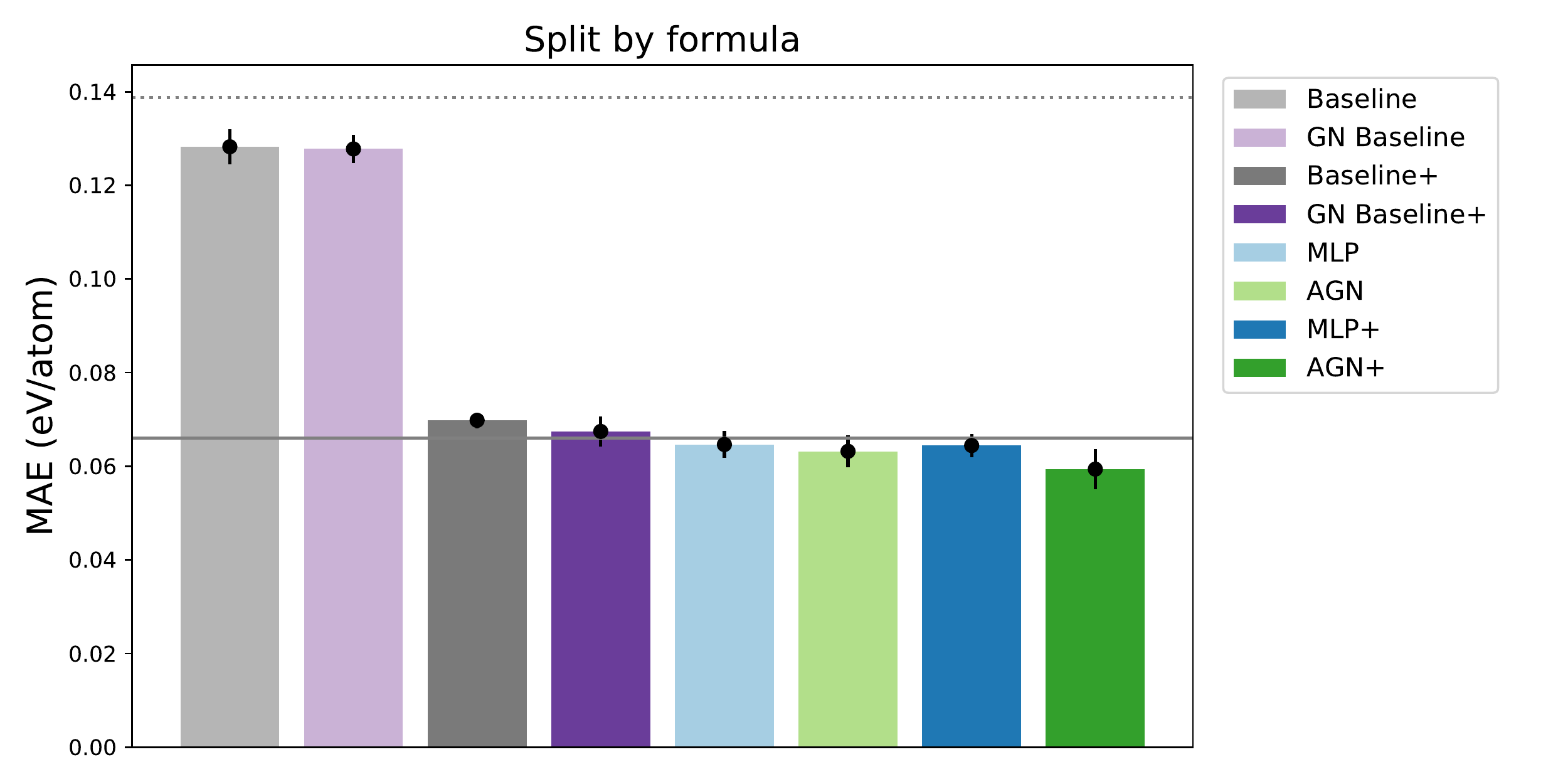}
    \caption{\tx{Mean absolute error on the experimental test set for the approach described in the text: the first four approaches do not perform transfer learning. The approaches denoted by a + use the DFT-computed formation energies as extra inputs. The Baseline and MLP approaches use an MLP as a network, while the GN Baseline and AGN approaches use a graph network.}}
    \label{fig:app_no_transfer}
\end{center}
\end{figure}

In \autoref{fig:app_no_transfer} we investigate the performance of our models when we ablate the transfer learning part. For the simplest MLP this corresponds to our baseline approach (modulo a minor difference in the network architecture). We denote by \emph{Baseline+} the approach where the model also receives the (simulated) DFT formation energies similarly to MLP+. \emph{GN Baseline} and \emph{GN Baseline +} are the corresponding graph network based approaches.

We observe that all approaches that have access to simulation data, either in the form of transfer learning or by being fed the simulated formation energies directly, perform significantly better than the baselines, regardless of the type of network used. Transfer learning performs better than simply getting the DFT computed target, a testimony of the importance of the features that can be learned during this process. However, the best approaches combine transfer learning with the ground truth targets and, as alluded to in the main text, are capable of learning significant corrections over it. In this case, the superior inductive bias of graph networks becomes more apparent.

% \paragraph{Influence of the simulation dataset}

% In \autoref{fig:app_simulation_ds} we investigate the effect  of changing the dataset which is used for training the simulation head of our model. In contrast to what was reported in Jha et al.\cite{Jha2019NatureComm}, we obtain better performance with the Materials Project data than with the OQMD data\cite{oqmd}. Note that some of our training parameters were not re-tuned when modifying the simulation dataset. We also investigate whether keeping only the most stable compound with a given stochiometric formula affects results (the graph networks based approach can distinguish between such crystal isomers but not the MLP based ones), but find that this has little to no effect on the results.

% \begin{figure}[h]
% \begin{center}
%     \includegraphics[width=\textwidth]{results_simulation_ds.pdf}
%     \caption{Mean absolute error on the experimental test set as a function of the training dataset used: Materials Project (MP) or OQMD. The variant indicated by a star denots the dataset where compounds with the same formula are removed (keeping only the most stable one)}
%     \label{fig:app_simulation_ds}
% \end{center}
% \end{figure}

\paragraph{Extended model analysis}

\begin{figure}[h]
\begin{center}
    \includegraphics[width=\textwidth]{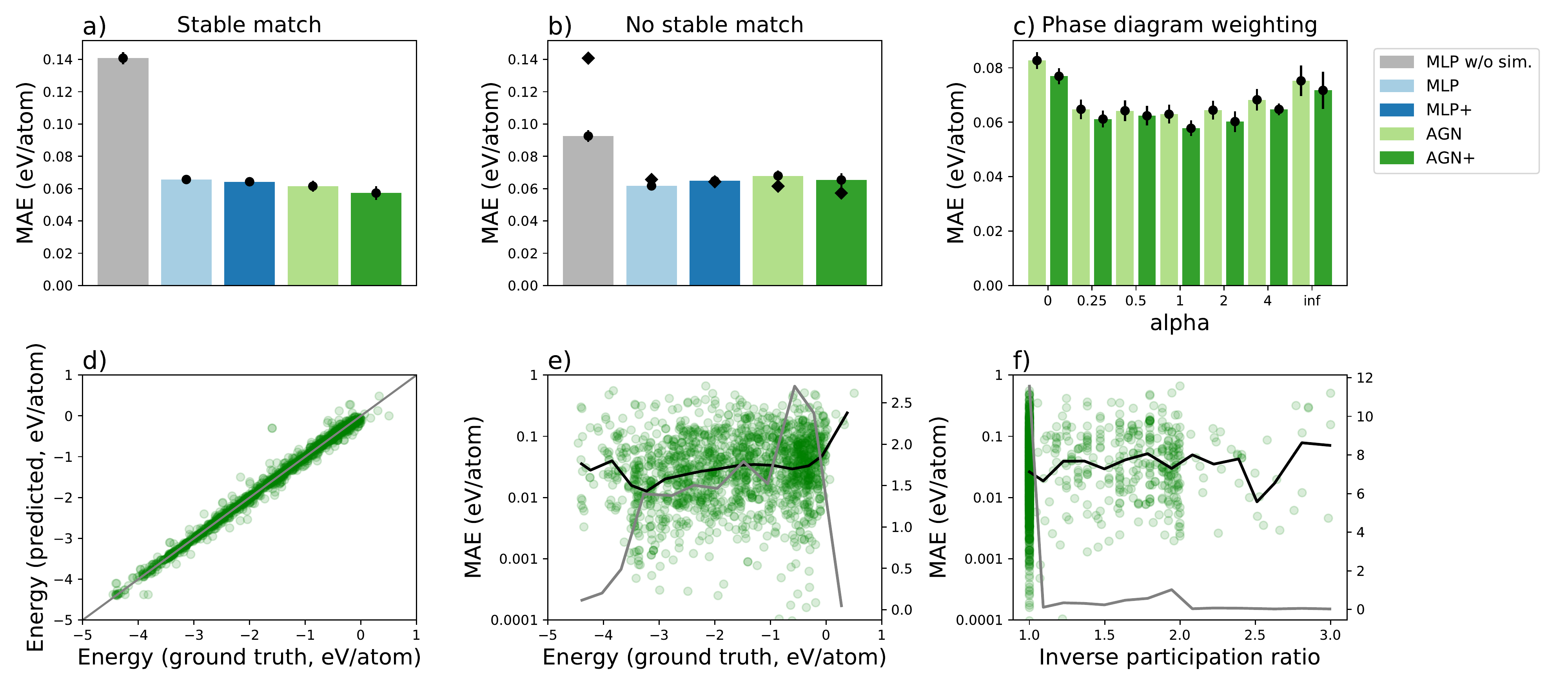}
    \caption{(a, b): mean absolute error on the test dataset, split by formula, depending on whether an experimental compound has a (resp. has no) stable match in the simulation dataset. The diamonds on the central panel are reminder of the values of the leftmost panel, highlighting the fact that the performance of the AGN approaches degrades when there is no stable match, while the MLP without simulation and MLP approaches improve. (c): mean absolute error as a function of the parameter $\alpha$ of Eq. (\ref{eq:gnn_update}). In both panels, each bar shows the average mean absolute error for 10 models trained with 10 different heldout validation split. Error bars are computed over the 10 different validation splits, for the 10 different models that we trained. (d-f): Scatter plot for the AGN+ model of (d) predicted formation energies against ground truth energies on the 10 validation splits, (e) mean absolute error against the formation energy, (f) mean absolute error against the inverse participation ratio (defined in the text). The black lines are binned averaged of the scattered points, while the grey line show the density of examples for a given value of the $x$-axis. }
    \label{fig:results_analysis_app}
\end{center}
\end{figure}

\autoref{fig:results_analysis_app}a and b provide supplementary information over Figure~\ref{fig:results_analysis_1}a, and confirm that all the non structured approaches perform better in the no stable match case while the structure approach always perform better when there is a stable match. \autoref{fig:results_analysis_app}c extends panel Figure~\ref{fig:results_analysis_1}c and reveals that the value of $\alpha=1$ is optimal for both the AGN and AGN+ methods.

\autoref{fig:results_analysis_app}d and e, investigate the predicted energies in more detail for the AGN+ model. We find that error (panel e) is relatively independent of the energy of the compound, although there is a slight increasing trend with the energy. In \autoref{fig:results_analysis_app}f, we plot the error against the inverse participation ratio $ipr(w) = \left(\sum_{i=1}^n w_i^2 \right)^{-1}$ -- a continuous measure of the number of neighbors in the phase diagram\footnote{For instance, if a compound is uniformly spread over $n$ neighbors, $w=(1/n, 1/n, \dots, 1/n)$, then $ipr(w)=n$.}. This provides a finer analysis of the error in the case where there is no stable match in the simulation  We see that error is relatively independent of the inverse participation ratio, with the main increase happening very near $ipr(w)=1$. This shows that the prediction difficulty increases more sharply when the crystal structure becomes unknown, but beyond that, the number of compounds in the phase diagram and the distance to the nearest compound does not have a strong influence.

\begin{figure}[h]
\begin{center}
    \includegraphics[width=0.8 \textwidth]{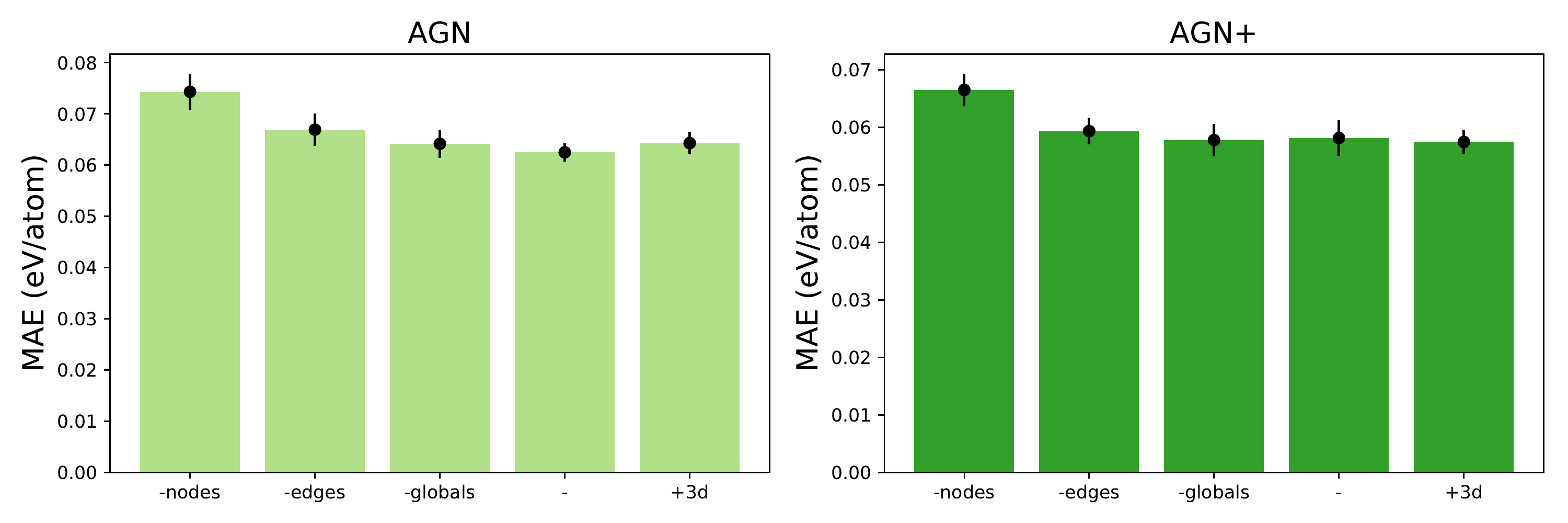}
    \caption{Performance of the AGN (\textbf{left}) and AGN+ (\textbf{right}) for various ablation conditions, described in the text.}
    \label{fig:ablation_results}
\end{center}
\end{figure}

\paragraph{Ablations of the graph network models} Finally we consider a series of ablation of our graph network models to unveil which features of the graphs are the most useful for the prediction. Ablating the global composition node from the graph by replacing it with a zero vector of the same shape (denoted "-globals" on \autoref{fig:ablation_results}) has the least severe effect, as the graph network can easily extract this information from the other nodes. The converse is not true, however, and wiping the atomic number from the nodes ("-nodes" on \autoref{fig:ablation_results}) degrades the performance more sizeably. Indeed, the nodes contain information not only about the global formula but also (when combined with the edges) about the distance patterns between atoms. Ablating the edges of the graph ("-edges", which contained distance information) only slightly deteriorates performance as well -- but at this stage it is important to remember that the graph that we feed to the network already contains \emph{some} distance information via the thresholding procedure which is used to connect vertices together or not. Finally, we experimented with feeding vector information rather than distances on the edges of the graph ("+3d" on \autoref{fig:ablation_results}). Because this quantity is not invariant to a global rotation of the crystal, in this case we perform a data augmentation under the form of a random 3d rotation of the network's input. Perhaps surprisingly, we find that this did not improve the results (it makes them slightly worse for the pure AGN model). We hypothesize that this may be related to subtle balance between performance on the simulation data and generalization to the experimental data of this machine learning setup.

\subsection{Bibliography}
\putbib[main]
\end{bibunit}

\end{document}